\documentclass[conference]{IEEEtran}
\IEEEoverridecommandlockouts
\usepackage{url}
\usepackage[hidelinks]{hyperref}
\usepackage{cite}
\usepackage{amsmath,amssymb,amsfonts}
\usepackage{algorithmic}
\usepackage{graphicx}
\usepackage{subcaption}
\usepackage{textcomp}
\usepackage{xcolor}
\usepackage{booktabs}
\usepackage[binary-units]{siunitx}
\usepackage{soul}
\usepackage{xcolor}
\usepackage{comment}
\usepackage[printonlyused,nohyperlinks, nolist]{acronym}

\def\BibTeX{{\rm B\kern-.05em{\sc i\kern-.025em b}\kern-.08em
    T\kern-.1667em\lower.7ex\hbox{E}\kern-.125emX}}
\begin{document}

\title{Towards an AI-enabled Connected Industry: AGV Communication and Sensor Measurement Datasets    \\
\thanks{This work was supported by the Federal Ministry of Education and Research (BMBF) of the Federal Republic of Germany as part of the AI4Mobile project (16KIS1170K). The authors alone are responsible for the content of the paper. Datasets are available at \url{https://dx.doi.org/10.21227/04ta-v128} \cite{04ta-v128-22}.}
\thanks{\copyright 2023 IEEE. Personal use of this material is permitted. Permission from IEEE must be obtained for all other uses, in any current or future media, including reprinting/republishing this material for advertising or promotional purposes, creating new collective works, for resale or redistribution to servers or lists, or reuse of any copyrighted component of this work in other works.}
}

\author{Rodrigo Hernang\'{o}mez\IEEEauthorrefmark{1}, \IEEEauthorblockN{Alexandros Palaios\IEEEauthorrefmark{4},  Cara Watermann\IEEEauthorrefmark{4},  Daniel Schäufele\IEEEauthorrefmark{1},\\ Philipp Geuer\IEEEauthorrefmark{4}, Rafail Ismayilov\IEEEauthorrefmark{1}, Mohammad Parvini\IEEEauthorrefmark{2}, Anton Krause\IEEEauthorrefmark{2},\\  Martin Kasparick\IEEEauthorrefmark{1},
Thomas Neugebauer\IEEEauthorrefmark{5}, Oscar D. Ramos-Cantor\IEEEauthorrefmark{3}, Hugues Tchouankem\IEEEauthorrefmark{3},\\  Jose Leon Calvo\IEEEauthorrefmark{4}, Bo Chen\IEEEauthorrefmark{7},
Gerhard Fettweis\IEEEauthorrefmark{2},
S{\l}awomir Sta\'{n}czak\IEEEauthorrefmark{1}\IEEEauthorrefmark{6}}

\IEEEauthorblockA{\IEEEauthorrefmark{1}Fraunhofer Heinrich Hertz Institute,  Germany, \{firstname.lastname\}@hhi.fraunhofer.de}

\IEEEauthorblockA{\IEEEauthorrefmark{4}Ericsson Research, Germany, \{alex.palaios, cara.watermann, philipp.geuer\}@ericsson.com}

\IEEEauthorblockA{\IEEEauthorrefmark{2}Vodafone Chair, Technische Universit\"{a}t Dresden, Germany, \{mohammad.parvini, anton.krause, gerhard.fettweis\}@tu-dresden.de}

\IEEEauthorblockA{\IEEEauthorrefmark{3}Corporate Research, Robert Bosch GmbH, Germany, \{oscardario.ramoscantor, huguesnarcisse.tchouankem\}@de.bosch.com}
\IEEEauthorblockA{\IEEEauthorrefmark{6}Network Information Theory Group,
Technische Universit\"{a}t Berlin, Germany}
\IEEEauthorblockA{\IEEEauthorrefmark{5}G\"{o}tting KG, Germany, neugebauer@goetting.de}
\IEEEauthorblockA{{\IEEEauthorrefmark{7}Enway GmbH, Germany, bo@enway.ai}}
}

\maketitle

\IEEEpubidadjcol

\begin{acronym}
    \acro{bosch}[iV2V]{industrial \acl{V2V}}
    \acro{enway}[iV2I+]{industrial \acl{V2I} plus Sensor}
    \acro{pQoS}{predictive Quality of Service}
    \acro{AI}{Artificial Intelligence}
    \acro{ML}{Machine Learning}
    \acro{CNN}{Convolutional Neural Network}
    \acro{LSTM}{Long Short Term Memory}
    \acro{UE}{User Equipment}
    \acro{RAN}{Radio Access Network}
    \acro{RRC}{Radio Resource Control}
    \acro{OFDM}{Orthogonal Frequency-Division Multiplexing}
    \acro{CE}{Consumer Equipment}
    \acro{DME}{Dedicated Measurement Equipment}
    \acro{GPS}{Global Positioning System}
    \acro{LAN}{Local Area Network}
    \acro{LOS}{Line-of-Sight}
    \acro{NLOS}{Non-Line-of-Sight}
    \acro{LTE}{Long Term Evolution}
    \acro{FDD}{Frequency-Division Duplex}
    \acro{RSRP}{Reference Signal Received Power}
    \acro{RSSI}{Received Signal Strength Indicator}
    \acro{RSRQ}{Reference Signal Received Quality}
    \acro{SNR}{Signal-to-Noise Ratio}
    \acro{RSCP}{Received Signal Code Power}
    \acro{EMM}{EPS Mobility Management}
    \acro{UL}{Uplink}
    \acro{DL}{Downlink}
    \acro{CSI}{Channel State Information}
    \acro{OAM}{Operation, Administration, Maintenance}
    \acro{BSR}{Buffer Status Report}
    \acro{ROC}{Receiver operating characteristic}
    \acro{SLA}{Service Level Agreement}
    \acro{NW}{Network}
    \acro{DMRS}{Demodulation Reference Signal}
    \acro{PSSCH}{Physical Sidelink Shared Channel}
    \acro{RFID}{Radio-frequency Identification}
    \acro{TDD}{Time Division Duplex}
    \acro{V2V}{Vehicle-to-vehicle}
    \acro{V2I}{Vehicle-to-infrastructure}
    \acro{V2X}{Vehicle-to-everything}
    \acro{KPI}{Key Performance Indicator}
    \acro{TP}{True Positive}
    \acro{FP}{False Positive}
    \acro{TN}{True Negative}
    \acro{FN}{False Negative}
    \acro{MMSE}{Minimum Mean Squared Error}
    \acro{LS}{least squares}
    \acro{AGV}{Automated Guided Vehicle}
    \acro{ROS}{Robot Operating System}
    \acro{EKF}{Extended Kalman Filter}
    \acro{LIDAR}{Light Detection and Ranging}
    \acro{IMU}{Inertial Measurement Unit}
    \acro{HD}{High Definition}
    \acro{3GPP}{3rd Generation Partnership Project}
    \acro{AGC}{Automatic Gain Control}
    \acro{QoS}{Quality of Service}
    \acro{QoE}{Quality of Experience}
    \acro{IoT}{Internet of Things}
    \acro{IIoT}{Industrial \aclu{IoT}}
    \acro{PoE}{Power over Ethernet}
    \acro{PPS}{Pulse-per-Second}
    \acro{SLAM}{Simultaneous Localization and Mapping}
    \acro{NTP}{Network Time Protocol}
    \acro{UDP}{User Datagram Protocol}
    \acro{SDR}{Software Defined Radio}
    \acro{HD}{High Definition}
    \acro{RUDE}{Real-time \acs{UDP} Data Emitter}
    \acro{CRUDE}{Collector for \acs{RUDE}}
    \acro{RRM}{Radio Resource Management}
    \acro{USRP}{Universal Software Radio Peripheral}
    \acro{MCS}{Modulation and Coding Scheme}
    \acro{QAM}{Quadrature Amplitude Modulation}
    \acro{SPS}{Semi-persistent Scheduling}
\end{acronym}
\begin{abstract}
This paper presents two wireless measurement campaigns in industrial testbeds: \acf{bosch} and \acf{enway}, together with detailed information about the two captured datasets. \ac{bosch} covers sidelink communication scenarios between \acp{AGV}, while \ac{enway} is conducted at an industrial setting where an autonomous cleaning robot is connected to a private cellular network. The combination of different communication technologies within a common measurement methodology provides insights that can be exploited by \ac{ML} for tasks such as fingerprinting, line-of-sight detection, prediction of quality of service or link selection.
Moreover, the datasets are publicly available, labelled and pre-filtered for fast on-boarding and applicability.

\end{abstract}

\acresetall

\section{Introduction}

Next generation wireless communication systems (5G and beyond) are expected to incorporate several new services and applications with demanding \ac{QoS} requirements, e.g., very low latency, minimum \ac{SNR}, delay, packet error rate, or very high \ac{UL} or \ac{DL} throughput. This challenge luckily benefits for
an increasing interest in \ac{pQoS}, i.e., \ac{QoS} estimation for a given time instance in the future. This can be done in different prediction horizons, ranging from milliseconds to hours or days.

\Acl{pQoS} can be particularly important for industrial wireless networks, where communication needs to be highly reliable due to, among other reasons, its integration into control loops. Wireless links are especially relevant in mobile setups, e.g., with one or more \acp{AGV} connected in a \ac{V2V}, \ac{V2I} or \ac{V2X} manner. In this regard, some datasets are available for automotive scenarios to train and test \ac{ML} algorithms and enhance such schemes\cite{schaufele2021terminal}. However, the availability of datasets from industrial and indoor measurement campaigns, such as \cite{burmeister2022_radioMap_DataSet}, is limited.
Proper knowledge of the \ac{QoS} conditions can ease industrial operation to guarantee human-machine safe interaction or robot cooperation. Other use cases include tele-operated driving, high-density platooning, and \ac{HD} map sharing for route selection \cite{kulzer2021ai4mobile,zorziAI6G}.

In this manner, we see a tendency towards the use of deep learning for \ac{pQoS} \cite{raca2020leveraging, minovski2021throughput, luo2018channel, ye2017power}. A consolidated overview of the \ac{ML}-enabled throughput prediction scenarios is presented in \cite{raca2020leveraging}. Likewise, \cite{minovski2021throughput} investigates a \ac{ML}-model to predict throughput in non-standalone 5G networks. 

\Ac{AI} and \ac{ML} have recently gained relevance in research yielding
an ample literature; however, \acs{AI}/\acs{ML} real-world implementation or validation remains elusive for industrial communication due to its high dependency on available datasets.
\Ac{AGV} use cases
share some unique characteristics, like a preferential use of private
campus networks in indoor radio environments and an emphasis
on (closed-loop) control~\cite{kulzer2021ai4mobile}, that are
untypical of automotive applications and other vehicular setups.
Therefore, creating a reference industrial dataset from experimental testbeds or simulations is paramount to evaluate fundamental theoretical models.

In this paper, we describe the \ac{bosch} and \ac{enway} datasets, which aim to pave the way for future experimentation of industrial mobile networks. The conducted measurement campaigns are part of a bigger data-collection framework covering also
automotive scenarios~\cite{palaios2021network, hernangomez2022berlin}. The common measurement procedure is described in detail in \cite{palaios2021network}
with some first results in \cite{palaios2021effect}.
The datasets are publicly available and documented in~\cite{04ta-v128-22}.

The remainder of this paper is structured as follows.  In Section \ref{bosch_full} and Section \ref{enway_full}, we describe the \ac{bosch} and \ac{enway} testbed and datasets and we elaborate on their details and components.
Section~\ref{studies} continues with an overview of potential \ac{ML} studies and Section \ref{conclusion} concludes the paper.

\begin{figure}[!t]
        \centering
        \includegraphics[width=\linewidth]{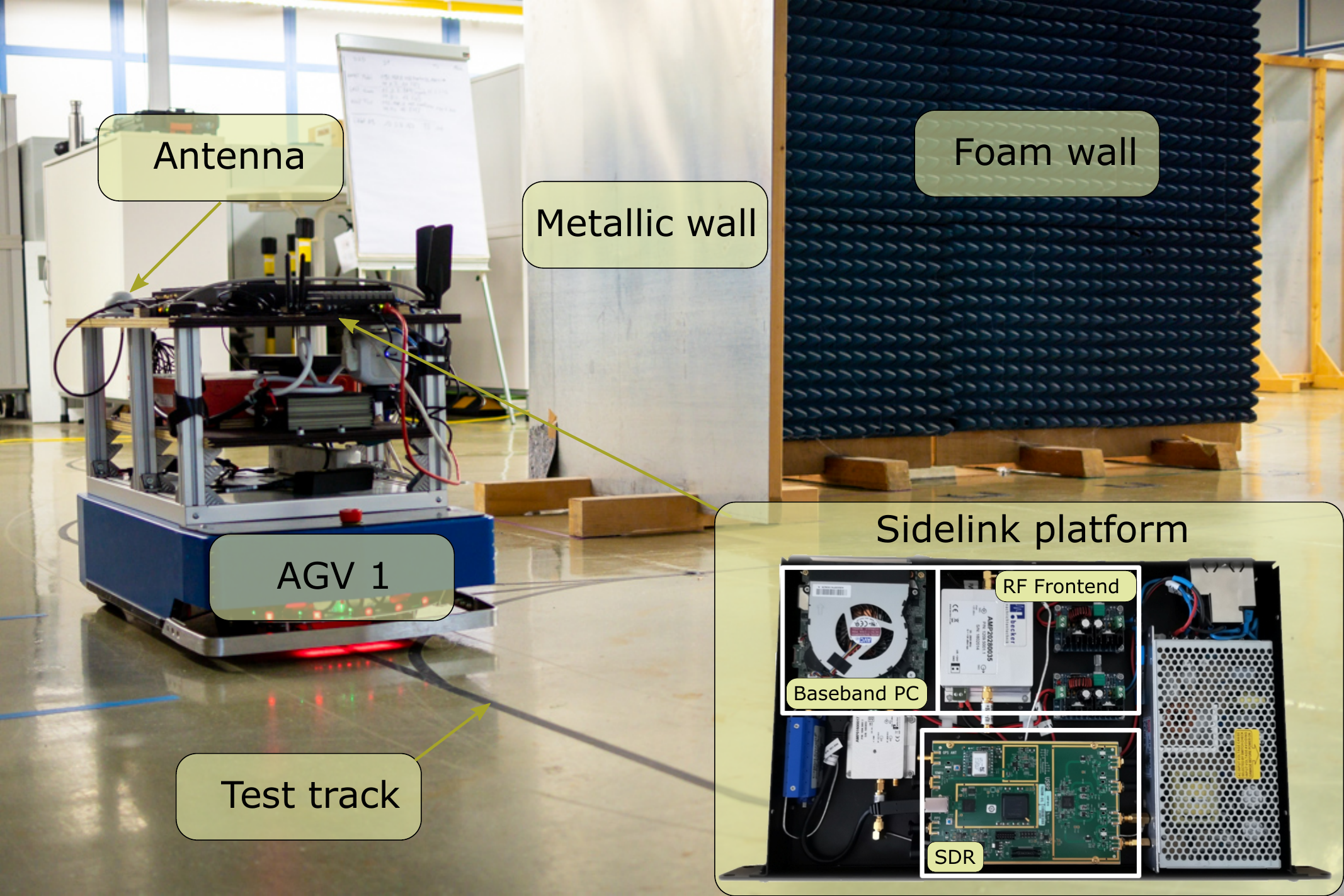}
    \caption{\ac{bosch} testbed.}
    \label{fig:bosch_testbed}
\end{figure}

\section{The \ac{bosch} Testbed and Dataset}\label{bosch_full}

The \ac{bosch} testbed entails three \acp{AGV} that exchange data with each other directly in a \ac{V2V} manner, using the sidelink technology as introduced by \ac{3GPP} in Release 14. One \ac{AGV} moves along a test track (cf. Fig.~\ref{fig:bosch_testbed}) while the other two \acp{AGV} remain static.

\subsection{Testbed Components}

\subsubsection{Sidelink}
\ac{3GPP} has standardized sidelink communication in 4G and 5G to define a 
\ac{V2V} communication
framework with varying interaction levels between devices and network.
Specifically,
mode 2 in 5G and mode 4 in 4G implement autonomous resource allocation, i.e.,
a decentralized procedure where 
\acp{UE} autonomously select resources and other transmit parameters~\cite{SL1}. For this, the sidelink operates in broadcast mode with \ac{SPS}.
While network-based resource allocation can outperform autonomous resource allocation due to the network's
orchestration role, the latter remains
the only option in the absence of network coverage.
 
For the measurement campaign, we use a full stack, software-based, standard-compliant and open implementation of the \ac{3GPP} Release 14 PC5 Mode 4 standard \cite{lindstedt2020}. The platform allows research concepts and standard features to be validated in hardware testbeds and it provides interfaces and tools for recording measurement data. Changes and adjustments are possible at every layer, which allows a realistic verification of new features. The sidelink software (all layers incl. baseband processing) can be run on standard general purpose computing hardware in connection with suitable \ac{SDR} hardware. We opted for a full stack implementation, thus providing a standard based IP to IP (one to all) interface for any application, i.e., all protocols on OSI layer 3 and higher can be transferred. The hardware setup is shown in Fig. \ref{fig:bosch_testbed} (cf. \cite{lindstedt2020} for further details). An omnidirectional antenna was connected to each \ac{SDR} platform and screwed onto the \ac{AGV}.

The \ac{SDR} platform was configured in the frequency band of \SI{3.7}{\giga\hertz} for both uplink and downlink with a bandwidth of \SI{10}{\mega\hertz}, 10 subchannels and a transmission gain of \SI{35}{\deci\bel}. A fixed \ac{MCS} of 12 was employed, which imposes the \ac{QAM} level 16. Together with an \ac{SPS} interval equal to \SI{20}{\milli\second}, the effective throughput reaches \SI{160}{\kilo\bit\per\second}.
 
\subsubsection{Localization}
Precise position of the communicating devices is required to link the environmental conditions with the measured data. For that purpose, the position information provided by the \acp{AGV} was recorded during the measurements. Two localization methods are used by different \acp{AGV} in the testbed: marker/track-based and \ac{SLAM}-based. In the former, the \acp{AGV} follow a track on the floor with an onboard camera. Additionally, \ac{RFID} tags are placed on the track to provide the exact position information when the \acp{AGV} pass over. Between \ac{RFID} tags, the \acp{AGV} estimate their position via odometry, i.e., based on wheel rotations, in combination with dead reckoning. Since the \acp{AGV} remain on track, the transversal error is below a few mm, while the longitudinal error was in the order of few cm.
In \ac{SLAM}-based localization, the \acp{AGV} use a laser scanner to detect and estimate the distance to landmarks, which are defined in the map of the \ac{AGV}. This method achieves a position accuracy in the order of few cm in our testbed.

The reported position of \ac{AGV} 1 was timestamped during the measurements.
Unless otherwise stated, the measurements scenarios presented below consider that the \ac{SLAM}-based \acp{AGV} were static and the marker/track-based \acp{AGV} were moving.

Since the moving \ac{AGV} 1 is guided by an optical line, the real lateral position is better than \textpm\SI{2}{\milli\meter} ($3\sigma$). The longitudinal error while passing an \ac{RFID} tag has a timing uncertainty of up to \SI{30}{\milli\second}, which gives an error up to \SI{30}{\milli\meter} at \SI{1}{\meter\per\second}. Dead reckoning results in additional errors due to a misalignment of the steering angle sensors to the route. The longitudinal error increases with the length of the unsupported route driven without an \ac{RFID} tag. The repeatability of the position information is below \textpm\SI{2}{\milli\meter} transversally and below \textpm\SI{2}{\centi\meter} longitudinally at the driven speed.

\subsubsection{Time Synchronization}\label{sec:time_sync}
To enable accurate evaluation of network latency and other \ac{QoS} properties, the time was synchronized across \acp{UE} via \ac{NTP} over Ethernet. The error is typically in the order of several \si{\micro\second} with a worst case error of \SI{1}{ms}.

\subsubsection{Controlled Packet Generation}
To ensure a precise periodic packet generation, we used a network packet generator tool based on \ac{RUDE} \& \ac{CRUDE} which is able to produce heterogeneous \ac{UDP} network traffic for realistic network workloads. It consists of two main modules: \ac{RUDE} generates traffic to the
network, which is then received and logged by the other module of the network with \ac{CRUDE}. We extended the packet generator tool to log all channel information for successfully received packets. The periodic packet transmission allows a simple way to estimate the packet error rate of the sidelink, similar as in~\cite{hernangomez2022berlin}.

\subsubsection{Automatic Gain Control}
The sidelink equipment also features \ac{AGC} based on \ac{RSSI} and \ac{RSRP} values
at the antenna input. \ac{AGC} is used to keep the received signal magnitude at a suitable level and is thus an elementary quantity for assessing \ac{QoS}.
\ac{AGC} calculations are already performed within the
\ac{SDR} platform so that the measured values supplied for pre-processing can be used as is.

\subsection{Measurement Scenarios}
\begin{figure}[!t]
    \includegraphics[width=\linewidth]{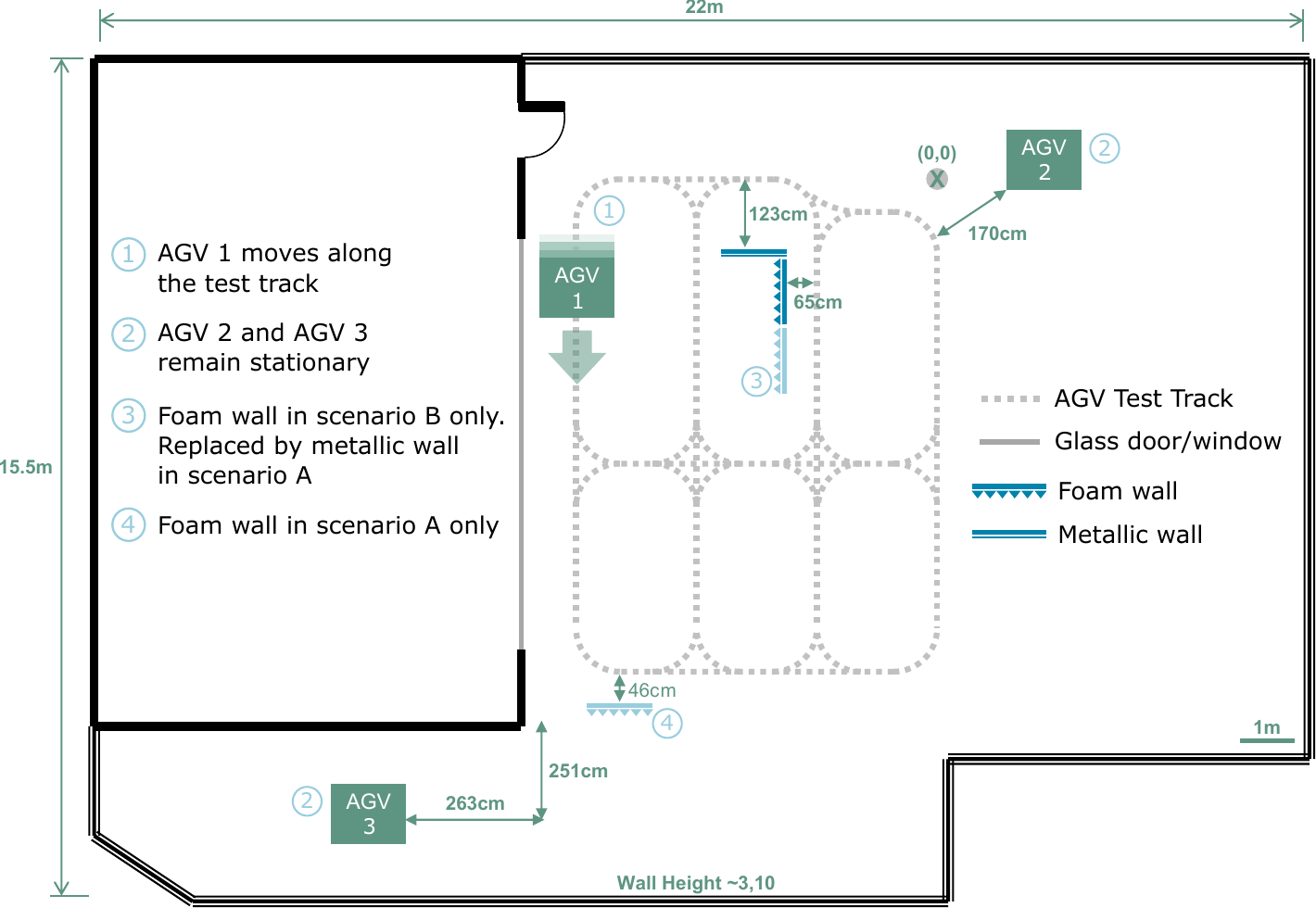}
  \caption{Illustration of measurement scenarios A \& B in \acs{bosch}.}
  \label{fig:measurement_scenario}
\end{figure}

We collected data for roughly 10 hours over the course of two days to acquire almost \SI{50}{\giga\byte} communication data between up to three industrial \acp{AGV}.
A schematic of the test area and its surroundings is presented in Fig. \ref{fig:measurement_scenario}. The dotted gray line depicts the track used by \ac{AGV} 1. Several obstacles, depicted in different blue tones in the figure, were located within the test area to achieve different radio propagation conditions, e.g., \ac{LOS} and \ac{NLOS} conditions. The obstacles were rearranged
during the measurement campaign to create two scenarios,
A and B, with different N/\ac{LOS} characteristics, as marked in light blue in the figure.

\subsection{The Dataset}
\subsubsection{Captured Sidelink Data}
For each scenario illustrated in Fig.~\ref{fig:measurement_scenario}, we capture the sidelink channel parameters for every transmitter/receiver pair.
A selection of captured sidelink channel parameters and their description are presented in Table~\ref{tab:sidelink_captured_data}.
The parameters in the table are obtained/estimated from the \ac{DMRS} of the \ac{PSSCH}.

\begin{table}[h]
\caption{Selected \ac{bosch} Data Features.}
\label{tab:sidelink_captured_data}
\resizebox{\columnwidth}{!}{%
\begin{tabular}{ll} \toprule
    {\textbf{Parameter}} & {\textbf{Description}}  \\ \midrule
    {SNR [dB]} & Derived from noise and power estimations of DMRS  \\
    {RSRP [dBm]}& Average energy per carrier/RE for DMRS  \\
    {RSSI [dBm]} & Signal power over the whole band  \\
    {Noise Power} & Estimated on DMRS in decoded subframe  \\
    {Time [sec]} & Receive time of first IQ-Sample of decoded subframe  \\
    {Frame Number} & System frame number  \\
    {Subframe Number}  & System subframe number \\
    {UHD Rx Gain [dB]} & Receive antenna gain  \\
    {SCI FRL N} & Starting subchannel of decoded PSSCH  \\
    {SCI FRL L} & Number of used subchannels for PSSCH  \\
    {RLC SN} & Sequence number of radio link control header  \\
    {Location} & Local x and y coordinates of \ac{AGV} 1 on the track  \\
     \bottomrule
\end{tabular}
}\vspace{3pt} \\

\end{table}
The \ac{AGV} localization data is provided as x and y coordinates in a local coordinate system.

\subsubsection{Dataset Pre-processing}\label{bosch:preprocess}
The dataset is constructed in a tabular format where each row represents a sample and the columns contain the value of the measured sidelink channel parameters.
Note that the sidelink parameters and \ac{AGV} location are measured independently and simultaneously in different devices.
With this setting, each measuring device embeds its own timestamp into the measured parameters.
In order to merge both measurements, we match a timestamped sidelink reading with a location measurement so that their timestamps differ by less than
\SI{5}{\milli\second}. Furthermore, we include labels for each sample indicating the measurement scenario (i.e., the obstacle placement) and the source and destination
\acp{AGV}.

\section{The \ac{enway} Testbed and dataset}\label{enway_full}
\subsection{Testbed Components}

\begin{figure*}[!t]
    \begin{subfigure}[t]{.45\textwidth}
        \centering
        \includegraphics[width=\linewidth]{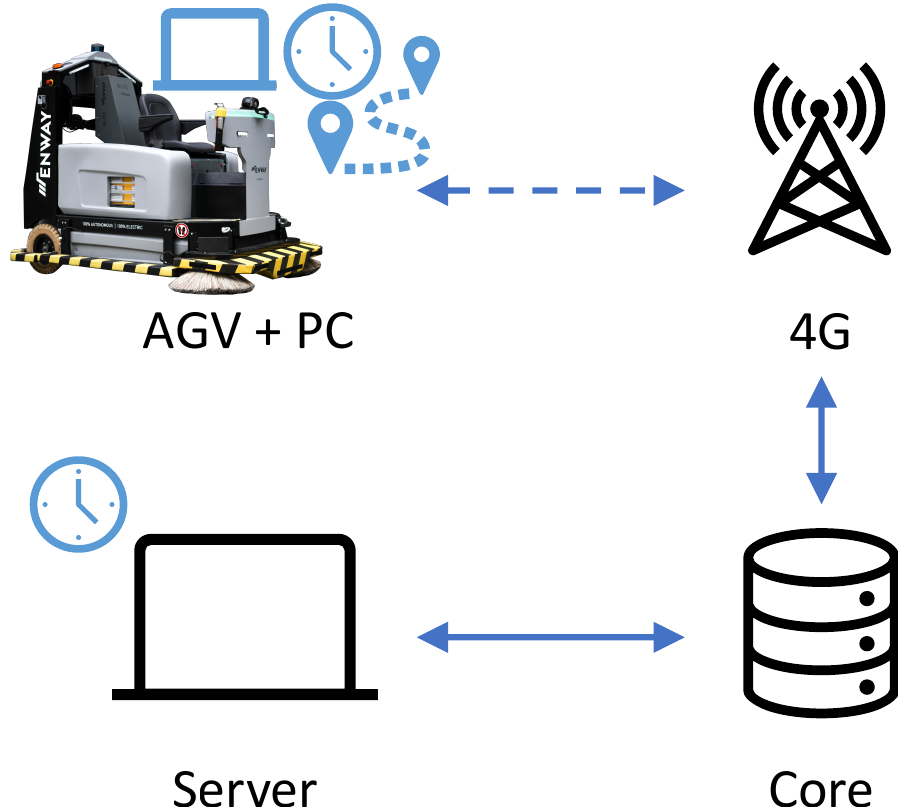}
        \caption{\ac{enway} setup. Wireless and cabled connections are represented with dotted and solid lines, respectively.
        }
        \label{fig:Arch}
    \end{subfigure}
    \hfill
    \begin{subfigure}[t]{.45\textwidth}
    \centering
        \includegraphics[width=\linewidth]{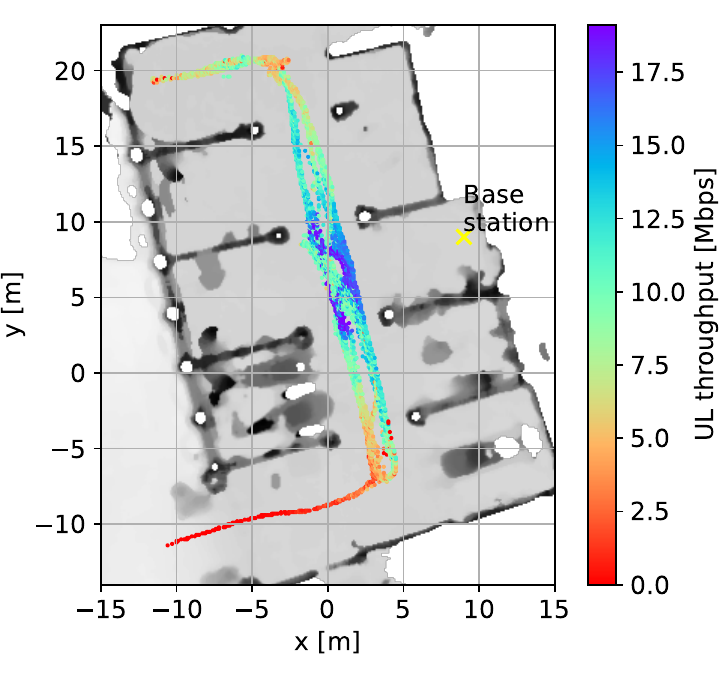}
        \caption{\acs{LIDAR}-based map of the environment with a throughput heatmap along the \ac{AGV}'s route. Walls appear in black.}
        \label{fig:Map}
    \end{subfigure}
    \caption{\ac{enway} testbed.}
    \label{fig:enway_testbed}
\end{figure*}

The testbed for the measurements is located in an industrial co-working space in Berlin, with a layout as shown in Fig.~\ref{fig:Map}. The hall had a gateway, which allowed the \ac{AGV} to drive outside. 

\subsubsection{The \ac{AGV}}
The \ac{AGV} used in the testbed is an autonomous cleaning robot from the company Enway, as shown in Fig.~\ref{fig:Arch}. They are specially designed for use under the operating conditions of the manufacturing industry.

The sweeper has all the necessary navigation data saved on a digital map, and drives over the cleaning area autonomously. Thanks to high-performance sensors and control software from Enway, the \ac{AGV} navigates the environment completely independently. Using a combination of laser distance measurement and cameras, the robot captures the environment in three dimensions. This 360-degree view enables very safe navigation between people, complex production lines, and overhanging systems. The \ac{AGV} immediately detects obstacles that suddenly appear along the route, and drives around them. Additional equipment such as floor markings, QR codes, or magnetic tracks are not required for navigation. In the event that the \ac{AGV} encounters an unsolvable situation, the robot stops and reports automatically to Enway headquarters via a data connection. The remote team monitors every movement of the device around the clock. The specialists can end autonomous journeys at any time, and can take control from a distance. Collisions with people, production systems, vehicles, and stored goods are thus avoided at all times. The machine can also be navigated manually by the operating personnel on site, if necessary. Because the \ac{AGV} is a retrofitted ride-on sweeper, it can be controlled from the driver's seat in the traditional sense. Ongoing use of the autonomous sweeper can be monitored, controlled, and then evaluated using mobile devices such as smartphones, laptops, or tablets. The software completely logs the cleaning trips.

\subsubsection{Cellular Network}
The mobile network used for the measurements in this testbed corresponded to a standardized 4G campus network with \ac{TDD} medium access. The bandwidth was \SI{20}{\mega\hertz} in the frequency band of \SIrange{3700}{3800}{\mega\hertz} approved by the Federal Network Agency. A corresponding frequency assignment was applied for the period of the measurements. 
The hardware consisted of a server running the \ac{LTE} core and a radio base station connected to the server via Gbit \ac{LAN} and powered via \ac{PoE}.
The base station exhibits an output power of
\SI{24}{\deci\bel\of{m}} and built-in antennas with a gain of \SI{6}{\deci\bel\of{i}} and an opening angle of \SI{60}{\degree} on both planes.
The location of the base station is marked in Fig. \ref{fig:Map} with a yellow cross.
A Mini-PC with the Linux operating system was placed on the \ac{AGV}'s seat to carry out \ac{QoS}-relevant measurements from a \ac{UE} perspective. A Quectel RM500Q-GL card was used as the radio device, which was connected to the Mini-PC via USB. External omni-directional antennas were attached above the seat back and connected to the radio.

The server provides a service interface to which the applications required for the measurements can be connected. The stationary applications can communicate
with mobile applications running in the Mini-PC 
via the service interface. This way, the location-dependent data rate and latency parameters relevant for evaluating the \ac{QoS} can be determined at the Mini-PC. 

\subsubsection{Time Synchronization}
The \ac{AGV} and the server were time synchronized. Fig.~\ref{fig:Arch} visualizes the communication setup. Both the server and the Mini-PC on the \ac{AGV} were connected to a GPS receiver with the sole purpose of time synchronization via \ac{PPS} signalling. The maximum error is typically in \si{\micro\second}-range. However, the \ac{AGV} only had consistent GPS reception at the start point and the outdoor area, so that the error can eventually accumulate up to the \si{\milli\second}-range.

\subsubsection{LTE Modem Access}
We used MobileInsight, an open-source cross-platform application for mobile network monitoring and analytics to capture mobile network data at the Mini-PC. It collects mobile network information across several cellular protocols, e.g. \ac{RRC} or \ac{EMM}. During the measurement campaign, the available information was logged every 40 ms.

Additionally, a Python script that accesses the modem via a virtual serial interface, a few radio parameters (\ac{RSSI}, \ac{RSRP}, \ac{RSRQ}, \ac{SNR}) were logged every 200 ms by the modem and written to a file with a time stamp. The data can then be linked to other data, e.g. the location, via the common time stamps.

\subsubsection{Controlled Packet Generation}
We used the application iperf3 on the Mini-PC and the server to generate \ac{UDP} traffic with a high or medium profile in either \ac{UL} or \ac{DL} direction. Apart from the generated iperf3 log files, tcpdump was used both on the Mini-PC and server to capture all incoming and outgoing packets at the respective network interfaces, allowing a more detailed evaluation on a packet level.

\subsection{Measurement Scenarios}
For the measurement campaign, one \ac{AGV}, equipped with the sensors and measurement devices described in the prior section, was driving through the testbed area over the course of three days. Overall, 16 hours of data were collected.

\ac{UL} and \ac{DL} communication was measured. Two types of packet flows were established to generate high and medium traffic profiles.
The throughput target for the high profile was
set to \SI{80}{\mega\bit\per\second} and
\SI{25}{\mega\bit\per\second} for
\ac{DL} and \ac{UL}, respectively.
In contrast to this,
the medium throughput target comprised
\SI{20}{\mega\bit\per\second} (\ac{DL}) and
\SI{10}{\mega\bit\per\second} (\ac{UL}).
Approximately twice as many high throughput measurements as medium throughput measurements were collected.
The route contained a diverse set of radio conditions, namely: \ac{LOS}  and \ac{NLOS} situations, coverage loss as well as indoor and outdoor measurements. The measured path is shown in Fig. \ref{fig:Map}.

\subsection{The Dataset}
For each scenario, data from the described network components and the sensors from the \ac{AGV} was collected. A subset of the captured data is presented in Table \ref{tab:enway_feat}.

\begin{table}[h]
\caption{Selected \ac{enway} Data Features.}
\label{tab:enway_feat}
\resizebox{\columnwidth}{!}{%
\begin{tabular}{ll} \toprule
    {\textbf{Parameter}} & {\textbf{Description}}  \\ \midrule
    {SNR [dB]} & Derived from noise and power estimations of DMRS  \\
    {RSRP [dBm]}& Average energy per carrier/RE for DMRS  \\
    {RSSI [dBm]} & Signal power over the whole band  \\
    {Throughput} & Acquired throughput in respective link direction \\
    {Ping [ms]} & Time in ms until a ping reply was received \\
    {Jitter} &  Delay variation measured over 1s \\
    {Odometry}        & Fused position, orientation and speed of the \ac{AGV}    \\
    {Map static elevation} & Single pre-computed map of the whole area \\
{Near/far map obstacles}    & \SI{36}{\meter\squared}/\SI{400}{\meter\squared} obstacle map around the \ac{AGV}     \\
\acs{LIDAR}                & 3D point cloud with obstacles \\
     \bottomrule
\end{tabular}
}
\end{table}

\subsubsection{\ac{AGV}-Sensor Data}

The \ac{AGV} delivers a series of sensor data via its \ac{ROS},
including the last fourth rows in Table~\ref{tab:enway_feat}. Except for \ac{LIDAR}, all \ac{ROS} topics shown here are obtained through sensor fusion
(e.g., through techniques such as \ac{EKF})
and provide relevant ground-truth information from a wireless perspective: position, orientation
and speed of the \ac{AGV} and location of walls and obstacles in different formats (2D, 3D, offline and online, near and far).
The raw input for the sensor fusion comes from sources including pure wheel odometry, drive commands, an \ac{IMU} and the already mentioned \ac{LIDAR}, all of them also available in the dataset.

\subsection{Pre-Processing}
Similar to what was described in Section \ref{bosch:preprocess} the data was pre-processed to further simplify the work with the collected data. Moreover, and since the position of the base station is known and fixed, the distance and clearance of the wireless link can be easily inferred from the sensor data and is provided as part of the pre-processed dataset.

We have merged the GPS timestamp logs, the LTE stack measurements and the throughput measurement together with the sensor-based link distance and link clearance into a single dataframe. As these data streams have different sampling frequencies, we re-sampled as needed to 1 second before the final merge. 

\section{Potential studies}\label{studies}

The measurement procedure of \ac{bosch} and \ac{enway} foremost pursues the study of \ac{pQoS}
in an industrial context,
as motivated by the existing literature~\cite{schaufele2021terminal,zorziAI6G,raca2020leveraging, minovski2021throughput, luo2018channel, ye2017power,palaios2021network,hernangomez2022berlin,palaios2021effect}.
Indeed, one can set diverse parameters as the target \ac{QoS}, such as delay, throughput, (for \ac{enway}, cf. Fig.~\ref{fig:Map})
or packet error rate (for \ac{bosch}). Physical layer parameters, such as \ac{SNR} or \ac{RSRP},
constitute the natural choice of input features for the prediction algorithms.
The localization data of both datasets can also be included as a valuable prediction input, as well
as the rich sensor data of \ac{enway} (cf. Table~\ref{tab:enway_feat}).

Beyond \ac{QoS} prediction, the presented datasets enable position-based tasks such as radio map estimation, fingerprinting and channel charting,
as well as the development of methods for proactive \ac{RRM}, like link selection.
Furthermore, the presence of fading conditions in the data allows the investigation of outlier detection, outage prediction, and other methods to increase the
resilience of industrial wireless networks.

\ac{ML} is a very powerful tool to tackle these problems,
and the datasets will help to investigate and demonstrate its benefits
over traditional handcrafted approaches in these areas. Finally, the variety of scenarios that we
covered in the two measurement campaigns offer the possibility of
studies around meta-learning.
Such techniques, which include transfer learning, domain adaptation or few-shot learning, can be explored with either dataset on its own or in combination with each other or with another related dataset~\cite{burmeister2022_radioMap_DataSet,hernangomez2022berlin}.

\section{Conclusion}\label{conclusion}

In this paper, we have described
\acf{bosch} and \acf{enway},
two testbeds for wireless communications in industrial settings.
We have provided detailed information about the components of the testbeds, together with the initial concept, the captured scenarios and some hints at possible future studies. The described datasets are publicly available for \ac{ML} research alongside documentation and code~\cite{04ta-v128-22},
what we consider a valuable contribution to the available industrial datasets both in terms of size and quality.

\ac{bosch} and \ac{enway} contain rich and complete data that we believe to be highly useful to answer questions regarding the use and generalisation of \ac{ML} for mobile use cases in industrial environments.
Both datasets represent realistic examples of their own kind in
a manageable size, which can enable the development of \acs{AI}-methods in a controlled manner as a first milestone and motivation towards exhaustive data collection procedures and large-scale deployments.

Thus, these datasets can be used to train and evaluate methods for \ac{pQoS}, which is a crucial enabler for high reliability in wireless industrial applications. Moreover, \ac{pQoS} in general, and our data in particular can be used as an ingredient for \ac{ML} algorithms that optimize the network itself, e.g., by performing proactive \ac{RRM}. This type of new network capabilities will be an important part of the future of wireless communications, such as the 6G cellular evolution. Finally, the addition of \ac{AGV} sensor data and localization opens the gate to advanced techniques like fingerprinting or channel charting.

\bibliographystyle{IEEEtran}
\bibliography{./bibliography.bib}
\vspace{12pt}

\section*{Biographies}

\begin{IEEEbiographynophoto}{Rodrigo Hernang\'{o}mez}
received his B.Sc. and M.Sc. in Telecommunication Engineering in 2015 and 2017 from Universidad Polit\'{e}cnica de Madrid, Spain. He is currently working towards his PhD as a research associate with the Fraunhofer HHI. His interests include machine learning and signal processing 
techniques for wireless communication and radar.
\end{IEEEbiographynophoto}
\vskip -2\baselineskip plus -1fil

\begin{IEEEbiographynophoto}{Alexandros Palaios}
received the doctorate degree from the RWTH Aachen university in 2016. He also received two Master's degrees in basic and applied cognitive science and in communication systems and networks from the University of Athens in 2008 and 2010. His research interests include data analysis and ML for wireless communications towards 6G.
\end{IEEEbiographynophoto}
\vskip -2\baselineskip plus -1fil

\begin{IEEEbiographynophoto}{Cara Watermann}
is a researcher at Ericsson Research in Germany. She received the M.Sc. in Computer Engineering from the University of Duisburg-Essen in 2021. She is currently working in the areas ML for wireless communications. Her research interests include ML-based pQoS and explainability.
\end{IEEEbiographynophoto}
\vskip -2\baselineskip plus -1fil

\begin{IEEEbiographynophoto}{Daniel Schäufele}
received the B.Sc. degree in IT-Systems engineering from the Hasso-Plattner-Institute, University of Potsdam, Germany, in 2012 and the B.Sc. and M.Sc. degrees in computer engineering from Technical University of Berlin, Germany in 2015 and 2017.
He is currently a research associate with the Fraunhofer HHI. His  interests are applying machine learning algorithms to wireless communication systems.
\end{IEEEbiographynophoto}
\vskip -2\baselineskip plus -1fil

\begin{IEEEbiographynophoto}{Philipp Geuer} is a researcher at Ericsson Research in Germany. He studied Electrical Engineering at RWTH Aachen University and finished his master's degree in 2019. His research interests are in the area of wireless networks, focusing on ML for network resilience, proactive resource management and QoS prediction. 
\end{IEEEbiographynophoto}
\vskip -2\baselineskip plus -1fil

\begin{IEEEbiographynophoto}{Rafail Ismayilov}
received his B.S. (2011) in Informatics and Computing Technics from Azerbaijan State Oil Academy, Baku, Azerbaijan, and the M.S. (2019) in Internet Technologies and Information Systems from Göttingen University, Germany. He is currently working on his Ph.D. in artificial intelligence for wireless communications at Fraunhofer HHI, Berlin, Germany. \end{IEEEbiographynophoto}
\vskip -2\baselineskip plus -1fil

\begin{IEEEbiographynophoto}{Mohammad Parvini}
received the M.Sc. degree in communication systems from Tarbiat Modares University, Tehran, Iran, in 2021. He is currently pursuing his PhD at the Vodafone Chair for Mobile Communications Systems at TU Dresden.
\end{IEEEbiographynophoto}
\vskip -2\baselineskip plus -1fil

\begin{IEEEbiographynophoto}{Anton Krause}
Anton Krause received the  Dipl.-Ing. degree in electrical engineering in 2022 from the TU Dresden, Germany and is currently pursuing his PhD at the Vodafone Chair Mobile Communications Systems at TU Dresden.
\end{IEEEbiographynophoto}
\vskip -2\baselineskip plus -1fil

\begin{IEEEbiographynophoto}{Martin Kasparick}
received the Dipl.-Ing. and the Dr.-Ing. degrees from the Technische Universität Berlin, Germany in 2009 and 2015. Since 2016, he leads the Signal and Information Processing Group in the Wireless Communications department at Fraunhofer HHI.
\end{IEEEbiographynophoto}
\vskip -2\baselineskip plus -1fil

\begin{IEEEbiographynophoto}{Thomas Neugebauer}
received a Dipl.-Ing. degree in 1989 from the University of Hannover in the area of radio frequency technology.
He has been working the last 30 years with Götting KG in industrial wireless communications and localisation. There, he is department leader in sensor development for AGV localisation and guidance and is responsible for the company's campus network. 
\end{IEEEbiographynophoto}
\vskip -2\baselineskip plus -1fil

\begin{IEEEbiographynophoto}{Oscar D. Ramos-Cantor}
received the M.Sc. and Dr. Ing. Degrees in electrical engineering from the Technical University of Darmstadt, Darmstadt, Germany, in 2012 and 2017. He joined Robert Bosch GmbH in 2018, where he investigates connected mobility solutions in automotive and industrial domains. His research interests include next generation mobile communications, Industry 4.0, and campus networks.
\end{IEEEbiographynophoto}
\vskip -2\baselineskip plus -1fil

\begin{IEEEbiographynophoto}{Hugues Tchouankem}
Hugues Tchouankem received the Dipl.-Ing. degree in electrical engineering and information technology from the RWTH Aachen, Germany, and the Ph.D. degree in electrical engineering and information technology from the University of Hannover, Germany. 
He is currently a senior research engineer at the Corporate Research Department of Robert Bosch GmbH
\end{IEEEbiographynophoto}
\vskip -2\baselineskip plus -1fil

\begin{IEEEbiographynophoto}{Jose Leon Calvo}
holds an M.Sc. (2014) and Ph.D (2018) in electrical engineering from Universidad Carlos III Madrid, and RWTH Aachen University, respectively. He joined Ericsson Research in 2018, where he has been involved in European and German research projects for 5G and 6G development. 
\end{IEEEbiographynophoto}
\vskip -2\baselineskip plus -1fil

\begin{IEEEbiographynophoto}{Bo Chen}
received his Masters degree in computer science from the University of Zurich, Switzerland. He’s worked in management consulting in various industries and is currently CTO and managing Director at Enway GmbH in Berlin, Germany. Enway focuses on autonomous sweeping solutions for industrial and municipal use.
\end{IEEEbiographynophoto}
\vskip -2\baselineskip plus -1fil

\begin{IEEEbiographynophoto}{Gerhard P. Fettweis }
Gerhard P. Fettweis [f'09] earned a Ph.D. at RWTH Aachen in 1990. After a postdoc at IBM Research, San Jose, CA, he joined TCSI, Berkeley, CA. Since 1994, he is Vodafone Chair Professor at TU Dresden. Since 2018 he is founding director of the Barkhausen Institute.
\end{IEEEbiographynophoto}
\vskip -2\baselineskip plus -1fil

\begin{IEEEbiographynophoto}{S{\l}awomir Sta\'nczak}
studied electrical engineering with specialization in control theory at the Wroc{\l}aw University of Technology and at the Technical University of Berlin (TU Berlin).
He received the Dipl.-Ing. degree in 1998 and the Dr.-Ing. degree
in electrical engineering in 2003, both from TU Berlin; the Habilitation degree
followed in 2006.
Since 2015, he has been a Full Professor for Network Information Theory with the TU Berlin and the head of the Wireless Communications and Networks department at Fraunhofer Institute for Telecommunications, Heinrich Hertz Institute (HHI).
\end{IEEEbiographynophoto}
\vskip -2\baselineskip plus -1fil

\vskip -2\baselineskip plus -1fil

\end{document}